\documentclass[11pt,a4paper]{article}

\renewcommand{\baselinestretch}{1}

\usepackage{fullpage}
\usepackage{titlesec}
\titleformat*{\section}{\large\bfseries}

\usepackage{latexsym,amssymb,amsthm,amsmath,amsfonts,graphicx,booktabs} 
\usepackage[round]{natbib}

\newcommand{\bx}{\mathbf{x}}

\newtheorem{theorem}{Theorem}[section]

\newtheorem{definition}[theorem]{Definition}

\usepackage{collab}
\collabAuthor{wm}{orange!85}{Werner}
\collabAuthor{dw}{magenta!85}{Dave}
\collabAuthor{ar}{blue!85}{Andreas}

\begin{document}

\begin{center}
\renewcommand{\baselinestretch}{1}
{ \large
{\bf Copula-based robust optimal block designs}
}

\vspace{0.5cm}

W.G.~M\"uller$^a$\footnote{Corresponding author: Department of Applied Statistics, Johannes Kepler University Linz, 4040 Linz, Austria; \texttt{werner.mueller@jku.at}}, A.~Rappold$^b$ and D.C.~Woods$^c$ \\[3ex]
$^a$ Department of Applied Statistics, Johannes Kepler University Linz, Austria \\[1ex]
$^b$ Plasser \& Theurer Connected G.m.b.H, Hagenberg im M\"uhlkreis, Austria \\[1ex]
$^c$ Southampton Statistical Sciences Research Institute, University of Southampton, UK
\end{center}

\renewcommand{\baselinestretch}{1}

\noindent Blocking is often used to reduce known variability in designed experiments by collecting together homogeneous experimental units. A common modelling assumption for such experiments is that responses from units within a block are dependent. Accounting for such dependencies in both the design of the experiment and the modelling of the resulting data when the response is not normally distributed can be challenging, particularly in terms of the computation required to find an optimal design. The application of copulas and marginal modelling provides a computationally efficient approach for estimating population-average treatment effects. Motivated by an experiment from materials testing, we develop and demonstrate designs with blocks of size two using copula models. Such designs are also important in applications ranging from microarray experiments to experiments on human eyes or limbs with naturally occurring blocks of size two. We present methodology for design selection, make comparisons to existing approaches in the literature and assess the robustness of the designs to modelling assumptions.  \\

\noindent Key words: Binary response; pseudo-Bayesian $D$-optimality; equivalence theorem; generalized linear model; marginal model.

\section{Introduction and motivation}\label{sec:intro}

Statistical design of experiments underpins much quantitative work in the biological, physical and engineering sciences, providing a principled approach to the efficient allocation of (typically sparse) experimental resources to address the aims of the study. Often, experiments aim to understand a process by modeling discrete data, for example arising from the observation of a binary or count response. For completely randomized experiments, assuming homogeneous experimental units, a generalized linear model (GLM) may provide an appropriate description and there has been much research into the construction of optimal and efficient designs for multi-factor GLMs, including \citet{wler}. \citet{ds06, ds08} and \citet{rwle}. See \citet{aw2015} for a comprehensive review.

 When heterogeneous experimental units can be grouped into more homogenous groups, or blocks, accounting for this grouping can improve the precision of inferences made from the experimental data. Methods to find block designs for discrete data have recently been proposed by, amongst others, \citet{woods+v_11}, \citet{niaparast-schwabe} and \citet{ww2015}. Two modeling paradigms have been adopted in the design literature: conditional models where the joint distribution of the data is derived by explicitly including block-specific random effects (e.g. generalized linear mixed models, \citealp{bres-clay}); and marginal models, where the dependence structure of the data is specified separately from the marginal distribution of each response (e.g. with parameters estimated via generalized estimating equations (GEEs), \citealp{liang-zeger}). For the linear model, these two modeling approaches coincide. In this paper, we find optimal designs under a marginal modeling approach when the intra-block dependence structure is defined via a copula. Such models are particularly appropriate when block effects are not of interest in themselves and the aim of the experiment is to understand the effects of treatment factors averaged across blocks. Optimal designs for marginal models using alternative definitions of the dependence structure have been found by \citet{jho}, \citet{ua2004} and \citet{vdvw2014}.
 
\begin{figure}[htb]
\begin{center}
\includegraphics[scale=.75]{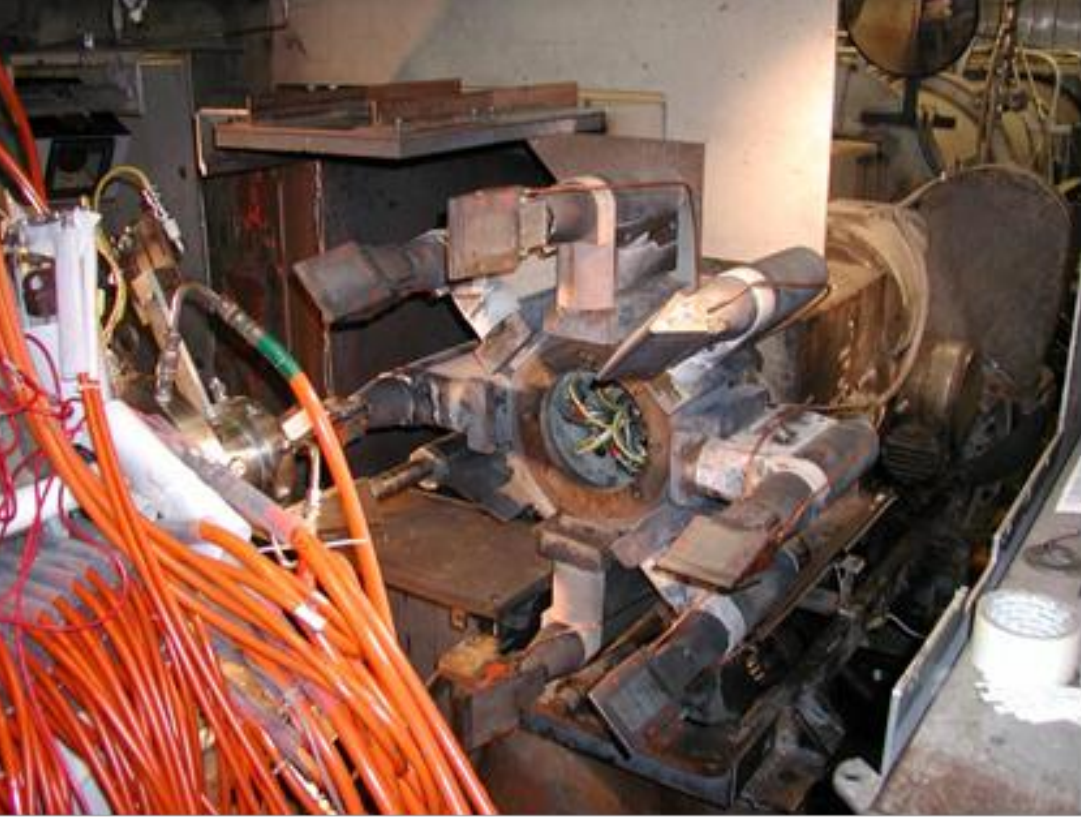} ~~~~~~
\includegraphics[scale=.75]{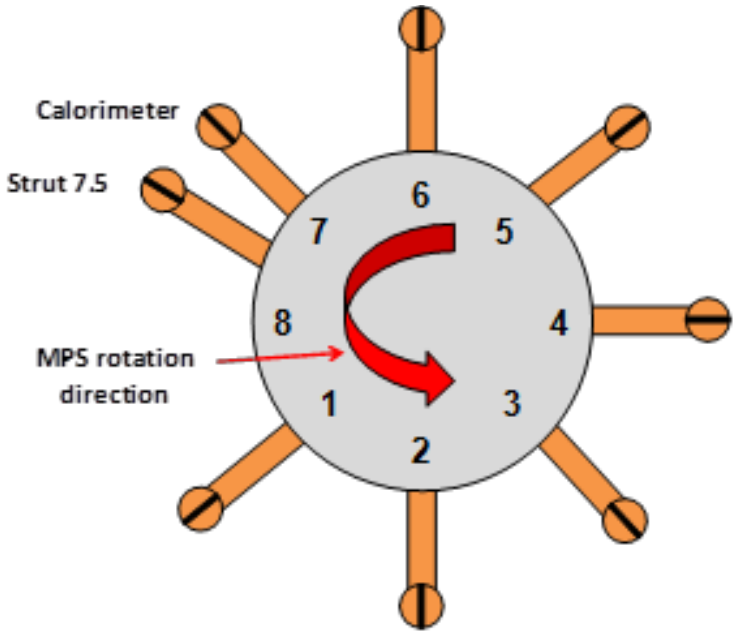}
\end{center}
\caption{\label{fig:arcjet}Arc jet carousel, struts and ``wedges'' (left) and schematic (right). In addition to the six wedges for holding material samples, the carousel had two further wedges used for temperature measurement.}
\end{figure}

 Although our methods can be generalized to arbitrary block sizes, we focus on the important special case of experiments with blocks of size two (see \citealp{godolphin2018}). Such blocks occur routinely in microarray experiments \citep{bailey_07, kerr_12} and in experiments on people, for example with eyes or arms as experimental units \citep{david+k_96}. Practical motivation for our work comes from a materials science experiment. In Section~\ref{sec:application} we find designs appropriate for aerospace materials testing experiments similar to those performed by our collaborators at the UK Defence Science and Technology Laboratory. The aim of these experiments is to compare the thermal properties of a set of novel materials against a reference material. In particular, one aim is to assess the probability of failure due to the exposure to extreme (high) temperatures. The experiment is performed using a arc jet to heat material samples which are held in one of six ``wedges'', each of which holds a pair of samples on a strut attached to a circular carousel, see Figure~\ref{fig:arcjet}. Hence, the experiment can be considered as a block design with six blocks, each containing two units. In the particular experiment considered here, six materials were tested, a reference and five novel samples. A variety of measures are made on each tested sample, including a visual inspection of quality to assess material failure which leads to a binary (pass/fail) response. It is this response for which we find optimal designs.

In common with most nonlinear models, the performance of a given design for a copula-based GLM model may depend on the values of the model parameters that define both the marginal model and the dependence structure. If strong prior information is available, then locally optimal designs can be sought for given values of the model parameters. Otherwise, Bayesian (e.g. \citealp{ow2017}) or maximin (e.g. \citealp{king+w:2000}) approaches can be adopted. In common with much of the recent literature on designs for GLMs, we find optimal designs robust to the values of the model parameters via a pseudo-Bayesian approach (e.g. \citealp{ADT2007}, ch.~18), with a classical quantity for design performance averaged with respect to a prior distribution on the parameters. Here, we adopt variants of $D$-optimality for design selection.     

The remainder of the paper is organized as follows. In Section~\ref{sec:copulas} we introduce the statistical models we employ, including copulas, and develop design methods for blocked experiments. An illustrative comparison is made to previous design approaches based on GEEs using an example from \citet{woods+v_11}. In Section~\ref{sec:application} we demonstrate and assess our methods via application to the materials testing example. In particular, we show how prior information on the parameters influences the choice of optimal design. We provide a brief discussion and some areas for future work in Section~\ref{sec:disc}.

\section{Designs for copula-based marginal models} \label{sec:copulas}
 
Suppose the experiment varies $m$ treatment factors, $\bx^T = (x_1,\ldots,x_m)$,  and the experiment has $b$ blocks of size $k$; throughout, our examples will assume $k=2$. The $j$th unit in the $i$th block receives treatment $\bx_{ij}^T=(x_{1ij},\ldots,x_{mij})$ $(i=1,\ldots,b;\, j=1,\ldots,k)$ and realizes observation $Y_{ij}$. The $\bx_{ij}$ are chosen from a standardized design space $\mathcal{X}=[-1,1]^m$ and are not necessarily distinct. Independence of observations $Y_{ij}, Y_{i'j'}$, for $i,i'=1,\ldots,b;\,j,j' =1,\ldots,k$, is assumed across blocks $(i\ne i')$ but we allow dependence within a block ($i=i'$), which we describe via a copula model.

\subsection{Statistical modeling via copulas}\label{sec:copmod}

The problem of specifying a probability model for dependent random variables $Y_{i1}, \dots, Y_{jk}$ can be simplified by expressing the corresponding $k$-dimensional joint distribution ${\mathbf{F}}_{{Y_{i1}},\dots,{Y_{ik}}}$ in terms of marginal distributions $F_{Y_{i1}}, \dots, F_{Y_{ik}}$, and an associated {$k$-copula} (or dependence function) $C$ defined as follows (cf. \citealp{nelsen_06}).

\begin{definition}
\label{Def:Copula}
A $k$-copula is a function $C: [0,1]^k \rightarrow [0,1]$, $k \geq 2$, with the following properties:
\begin{enumerate}
\item (\emph{uniform margins}) for every $\mathbf{u} \in [0,1]^k$, if at least one coordinate of $\mathbf{u}$ is $0$, then
\[C(\mathbf{u}) = 0\,, \]
and if all coordinates of $\mathbf{u}$ are $1$ except $u_i$, then
$$C(\mathbf{u}) = u_i\,.$$ 
\item (\emph{k-increasing}) for all $\mathbf{a}$, $\mathbf{b} \in [0,1]^k$ such that $\mathbf{a}\leq \mathbf{b}$,
\[V_{C}([\mathbf{a},\mathbf{b}]) \geq 0,\]
where $V_{C}$ is the measure induced by $C$ on $[0,1]^k$.
\end{enumerate}
\end{definition}

The connection between a copula and a joint probability distribution is given by Sklar's Theorem \citep{sklar_59}, which affirms that for every $k$-dimensional joint distribution ${\mathbf{F}}_{{Y_{i1}},\ldots,{Y_{ik}}}$ with marginal distributions $F_{Y_{i1}}, \ldots, F_{Y_{ik}}$, there exists a $k$-copula $C$, defined as in Definition~\ref{Def:Copula}, such that
\begin{equation}\label{Eq:S}
\mathbf{F}_{Y_{i1},\ldots,Y_{ik}} (y_1,\dots,y_k) = C(F_{Y_{i1}}(y_1),\ldots, F_{Y_{ik}}(y_k))\,,
\end{equation}
for all $y_1, \ldots, y_k\in\mathbb{R}$.
%
Conversely, if $C$ is a $k$-copula and $F_{Y_1}, \dots, F_{Y_k}$ are distribution functions, then the function $F_{Y_1,\dots,Y_k}$ given by (\ref{Eq:S}) is a joint distribution with marginals $F_{Y_1}, \dots,F_{Y_k}$. The copula $C$ may not be unique for discrete margins, however the practical limitations for statistical purposes are little, cf. \cite{genest+n_07}.

Owing to Sklar's theorem, parametric families of copulas represent a powerful tool to describe the joint relationship between dependent random variables. Selecting the appropriate dependence within an assumed parametric copula family reduces to the selection of copula parameters, which correspond, for example, to a specific measure of association for the modeled random variables. Assuming $Y_{i1},\ldots,Y_{ik}$ are continuous random variables with associated copula $C(\cdot;\alpha)$, one measure of association proposed by \cite{joe_90} is given by
\begin{equation}
\label{eq:tau}
\tau_k = \frac{1}{2^{k-1}-1} \left\{2^k \int\limits_{[0,1]^k} C(\cdot;\alpha) d C(\cdot;\alpha) - 1 \right\}\,.
\end{equation}
Equation~\eqref{eq:tau} is a generalized version of Kendall's $\tau$, and hence establishes a correspondence between a scalar copula parameter $\alpha$ and the degree of dependence. More details and properties of this quantity, and another more traditional measure of concordance, can be found in \citet{genest+al_11}.

\subsection{Design of experiments for copula models}\label{sec:design}

In common with most work on optimal design of experiments, we base our criterion on the Fisher information matrix (FIM), the inverse of which provides an asymptotic approximation to the variance-covariance matrix of the maximum likelihood estimators of the model parameters. 

Let $\zeta_i = (\mathbf{x}_{i1},\ldots,\mathbf{x}_{ik})\in\mathcal{X}^k$ denote the $k$ treatment vectors assigned to the units in block $i$ $(i = 1,\ldots, b;\,j = 1, \ldots, k)$.
We will work within a class of normalized block designs defined as 
$$
\xi = \left\{
\begin{array}{ccc}
\zeta_1,& \ldots, & \zeta_n \\
w_1, & \ldots, & w_n
\end{array}
\right\}\,,
\quad 0< w_i \leq 1\,,
\quad \sum_{i=1}^nw_i = 1\,,
$$
with $n \leq b$ distinct (support) blocks. As defined, $bw_i$ must be integer and represents the replication of the $i$th support block ($i=1,\ldots,n$). Without loss of generality, we assume the first $n$ blocks in the design correspond to $\zeta_1,\ldots,\zeta_b$, with the remaining $b-n$ blocks being replicates. We relax the assumption that $bw_i$ is integer to find so-called approximate or continuous designs; see also \citet{cheng} and \citet{ww2015}. Let $\Xi$ denote the space of all possible designs of this form.

Denote the vector of responses from the $i$th block as
$$
\mathbf{Y}_i = \left(Y_{i1},\ldots, Y_{ik}\right)^T\,,\quad i = 1,\ldots, b\,,
$$
with corresponding expectation vector
$$
\boldsymbol{\eta}_i = \left[\eta(\mathbf{x}_{i1};\,\boldsymbol{\beta}),\ldots,\eta(\mathbf{x}_{ik};\,\boldsymbol{\beta})\right]^T\,,
$$
where $\eta(\cdot;\,\cdot)$ is a known function and $\boldsymbol{\beta}=(\beta_1, \ldots,\beta_r)^T$ is a vector of unknown parameters requiring estimation. Denote the marginal distribution function for the $j$th entry in the block as $F_{Y_{ij}}\left(y_{ij};\, \mathbf{x}_{ij}, \boldsymbol{\beta}\right)$, $j=1,\ldots, k$, and denote the joint distribution, derived via a copula transformation, for the $k$ responses in the $i$th block as $C\left(F_{Y_{i1}},\ldots,F_{Y_{ik}};\, \boldsymbol{\alpha}\right)$ where $\boldsymbol{\alpha}=({\alpha}_1,\ldots, {\alpha}_l)^T$ are unknown (copula) parameters.

The FIM $M(\zeta_i;\,\boldsymbol{\gamma})$ for the $i$th block is an $(r +l) \times (r +l)$ matrix with $vw$th element  
\begin{equation}\label{Eq:FIM}
M(\zeta_i;\boldsymbol{\gamma})_{vw} = \mathbf{E} \left(  - \dfrac{\partial^2}{\partial \gamma_v \partial \gamma_w} \log c_{\mathbf{Y}_i}(\boldsymbol{\eta}_i, \boldsymbol{\alpha}) \right)\,,
 \end{equation}
where $\boldsymbol{\gamma}=(\gamma_1,\ldots,{\gamma}_{r+l})^T=({\beta}_1,\ldots,{\beta}_r,{\alpha}_1,, \ldots, {\alpha}_l)^T$ and 
\[
 c_{\mathbf{Y}_i}(\boldsymbol{\eta}_i, \boldsymbol{\alpha})= \dfrac{\partial^k}{\partial y_{i1} \dots \partial y_{ik}} C\left(F_{Y_{i1}},\ldots,F_{Y_{ik}};\, \boldsymbol{\alpha}\right)\]
is the joint density function represented through a copula $C$ in accordance with Equation~(\ref{Eq:S}). The FIM for an approximate block design $\xi$ is then given by
\[M(\xi;\, \boldsymbol{\gamma}) = \sum\limits_{i=1}^n w_i M(\zeta_i;\,\boldsymbol{\gamma})\,.\] 


An optimal design $\xi^\star$ maximizes a scalar function $\psi\left\{M(\xi;\,\boldsymbol{\gamma})\right\}$ of the information matrix. Previous work on optimal designs for copulas has focussed on finding completely randomized locally-optimal designs for multivariate responses, which can be considered as a block design where every unit within a block must receive the same treatment. \citet{denman_design_2011} found $D$-optimal designs for a bivariate response ($k=2$) that maximized $\psi^D \left\{M(\xi;\,\boldsymbol{\gamma})\right\} = \det M(\xi;\,\boldsymbol{\gamma})$, and \citet{Perrone+m_16} developed a corresponding equivalence theorem. These methods were extended to the local $D_A$-criterion, and, as a special case, for the $D_s$-criterion in \citet{perrone+al_17}. Other relevant uses of design of experiments in copula models are \citet{deldossi_optimal_2018} and \citet{durante_asymmetric_2016}, but until now all relied on the availability of a single ``best guess'' vector of parameter values.

To overcome this dependence on assumed parameter values, here we adopt a pseudo-Bayesian approach to constructing block designs. Furthermore, our primary interest is typically in $s$ meaningful linear combination of the parameters. Such combinations can be defined as elements of the vector $A^T\boldsymbol{\gamma}$, where $A^T$ is an $s \times (r+l)$ matrix of rank $s < (r+l)$. If $M(\xi;\, \boldsymbol{\gamma})$ is non-singular, the variance-covariance matrix of the maximum likelihood estimator of $A^T\boldsymbol{\gamma}$ is proportional to $A^T \{ M(\xi;\, \boldsymbol{\gamma}) \}^{-1} A$. Hence, we define a \textit{robust $D_A$-optimal block design} $\xi^\star$ as the design that maximizes 
\begin{equation}\label{eq:bayesDA}
\Psi^D(\xi;\,G, A) = \int_{\Gamma} \log \det[A^T \{ M(\xi;\, \boldsymbol{\gamma}) \}^{-1} A]^{-1}\,\mathrm{d}G(\boldsymbol{\gamma})\,,
\end{equation} 
\noindent where $G(\boldsymbol{\gamma})$ is a proper prior distribution function for $\boldsymbol{\gamma}$ and $\Gamma\subset\mathbb{R}^{r+l}$ is the support of $G$. See also \citet{woods+v_11}. 

Most often the main interest is in an $s < (r+l)$-dimensional subset of the parameters. In such a case, a \textit{robust $D_s$-optimal block design} can be found by maximizing 
\begin{equation}\label{eq:bayesDs}
\Psi^D(\xi;\,G) = \int_{\Gamma} \log \det \left\{M_{11} - M_{12}M_{22}^{-1}M_{12}^T\right\}\,\mathrm{d}G(\boldsymbol{\gamma})\,, 
\end{equation}
following the partition of the information matrix as
$$M(\xi;\, \boldsymbol{\gamma}) = \left (
\begin{array}{cc}
M_{11} &  M_{12} \\
M_{12}^T &  M_{22}
\end{array}
\right )\,.$$
Here, $M_{11}$ is the $(s \times s)$ partition related to the parameters of interest. This criterion follows as a special case of the $D_A$-criterion with $A^T = (I_s \; 0_{s\times (r+l-s)})$, with $I_s$ the $s\times s$ identity matrix and $0_{s\times (r+l-s)}$ the $s\times (r+l-s)$ zero matrix.


We evaluate a design $\xi$ via its \textit{Bayesian efficiencies} under a given criterion, relative to an appropriate reference design $\xi^*$ (see, for example, \citealp{waite2018}). Under robust $D_s$-optimality, this efficiency is given by:
\[
\text{eff}(\xi,\xi^*) =
\left(\dfrac{\exp\int_{\mathcal{B}} \log \det[M_{11}(\xi, {\boldsymbol{\gamma}}) - M_{12}(\xi, {\boldsymbol{\gamma}})M_{22}^{-1}(\xi, {\boldsymbol{\gamma}})M_{12}^T(\xi, \tilde{\boldsymbol{\gamma}})] \,\mathrm{d}F(\boldsymbol{\gamma})}{\exp\int_{\mathcal{B}} \log \det[
M_{11}(\xi^*, {\boldsymbol{\gamma}}) - M_{12}(\xi^*, {\boldsymbol{\gamma}})M_{22}^{-1}(\xi^*,{\boldsymbol{\gamma}})M_{12}^T(\xi^*,{\boldsymbol{\gamma}}) ]\,\mathrm{d}F(\boldsymbol{\gamma}) }\right)^{1/s}.
\]

We find designs that maximize~\eqref{eq:bayesDA} and~\eqref{eq:bayesDs} numerically using a version of the Fedorov-Wynn algorithm \citep{wynn,fedorov}, as implemented in \texttt{R} package \texttt{docopulae} \citep{docopulae}.

The optimality of a block design $\xi^\star$ under the robust $D_A$-criterion, regardless of how it was found, can be assessed via application of the following Kiefer-Wolfowitz-type equivalence theorem. The proof is similar to that for completely randomized experiments with multivariate response, see \citet{perrone+al_17} for the locally-optimal design case.
 
\begin{theorem}\label{Th:1}
The following properties are equivalent:
\begin{enumerate}
 \item $\xi^\star$ is $D_A$-optimal;
\item for every $\zeta \in \mathcal{X}^k$,
$$\int_{\mathcal{B}}
\textnormal{ tr }[ M(\xi^\star;\, {\boldsymbol{\gamma}})^{-1} A (A^T M(\xi^\star;\, {\boldsymbol{\gamma}})^{-1} A)^{-1} A^T M(\xi^\star;\, {\boldsymbol{\gamma}})^{-1} M(\zeta;\, {\boldsymbol{\gamma}})]  \,\mathrm{d}G(\boldsymbol{\gamma})               \leq s\,;$$  
\item over all $\xi \in \Xi$, the design $\xi^\star$ minimizes the function
$$\max\limits_{\zeta \in \mathcal{X}^k}
\int_{\mathcal{B}} \textnormal{ tr }[M(\xi^\star, {\boldsymbol{\gamma}})^{-1} A (A^T M(\xi^\star, {\boldsymbol{\gamma}})^{-1} A)^{-1} A^T M(\xi^\star, {\boldsymbol{\gamma}})^{-1} M(\zeta;\, {\boldsymbol{\gamma}})] \,\mathrm{d}G(\boldsymbol{\gamma})\,,$$
\end{enumerate}
where $\Xi$ is the set of all possible block designs.
\end{theorem}

\subsection{Comparative example}

We demonstrate robust optimal block designs for copula models using a simple example from \citet{woods+v_11}, which allows comparison to the designs found by those authors for a GEE model. We find robust designs for a single-factor log-linear regression model assuming Poisson marginal distirbutions and quadratic linear predictor, implying $\log\{\eta(\mathbf{x};\,\boldsymbol{\beta})\} = \beta_0 + \beta_1x + \beta_2x^2$. The prior distribution $G$ is uniform on the parameter space $[-1,1]\times [4, 5] \times [0.5, 1.5]$. In line with our motivating example, we assume blocks of size $k=2$ and intra-block dependence defined according to one of the following bivariate copula functions.
\begin{enumerate}
\item \emph{Product Copula}, which represents the independence case,
\[C(u_1,u_2) = u_1 u_2\,,\] 
with generalized Kendall's $\tau$ of $\tau_2 = 0$.

\item  \emph{Clayton Copula},
\item[]  \[{C}_{\alpha}(u_1,u_2;\,\alpha) =\big[  \max\big( u_1^{-\alpha} + u_2^{-\alpha} -1 ,\, 0\big) \big]^{-\frac{1}{\alpha}}\,,\] 
    with $\alpha \in (0, +\infty)$ and generalized $\tau_2 = \frac{\alpha}{\alpha + 2}$.
\item \emph{Gumbel Copula},
\item[] \[{C}_{\alpha}(u_1,u_2;\,\alpha) =\exp \big( - \big[ ( -\ln u_1)^{\alpha} + (-\ln u_2)^{\alpha} \big]^{\frac{1}{\alpha}} \big)\,,\] 
with $\alpha \in [1, +\infty)$ and generalized $\tau_2 = \frac{\alpha - 1}{\alpha}$.
\end{enumerate}

The first copula is chosen for reference purposes; the latter two represent opposing dependencies in the tails (lower tail dependence for the Clayton versus upper tail dependence for the Gumbel). To isolate the effect of the copula structure from the strength of the dependence, we set $\alpha$ for each copula such that the values for Kendall's $\tau$ coincide at three level,s $\tau_2=\epsilon>0, 1/10, 1/3$ respectively. Here $\epsilon=10^{-9}$ is a small number  to approximate the zero case, but avoid singularity issues.

To find robust $D$-optimal designs, objective function~(\ref{eq:bayesDA}) was evaluated using quadrature \citep{gjs}. Optimal designs under the Clayton and Gumbel copulas are shown in Figure~\ref{fig:toyexample}, and demonstrate that increasing the generalized dependence (i.e. increasing $\tau_2$) leads to designs placing more weight on support blocks with points on the edge of the design space. All the designs display a ``mirror-image'' structure, with all design points having $\mathbf{x}>0$. These features are common in designs for Poisson regression (see \citealp{rwle}). The designs found under the Gumbel copula tend to include more support blocks but the pattern in the changes to these blocks as $\tau_2$ is increased is similar for both copulas.

\newcommand{\compscale}{.35}
\begin{figure}[htb]
\begin{center}
\includegraphics[viewport = 100 250 500 600, clip, scale = \compscale]{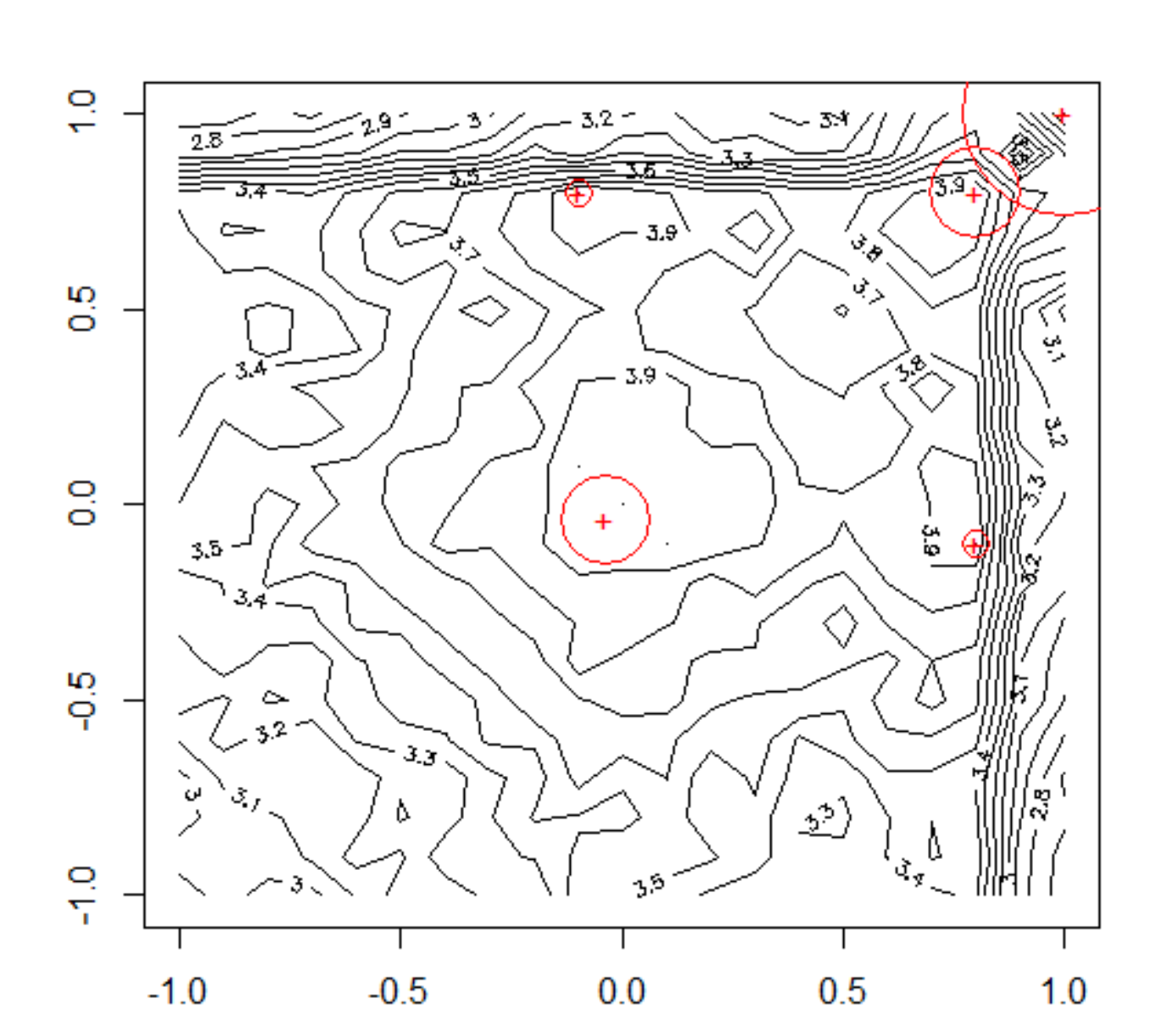}
\includegraphics[viewport = 100 250 500 600, clip, scale = \compscale]{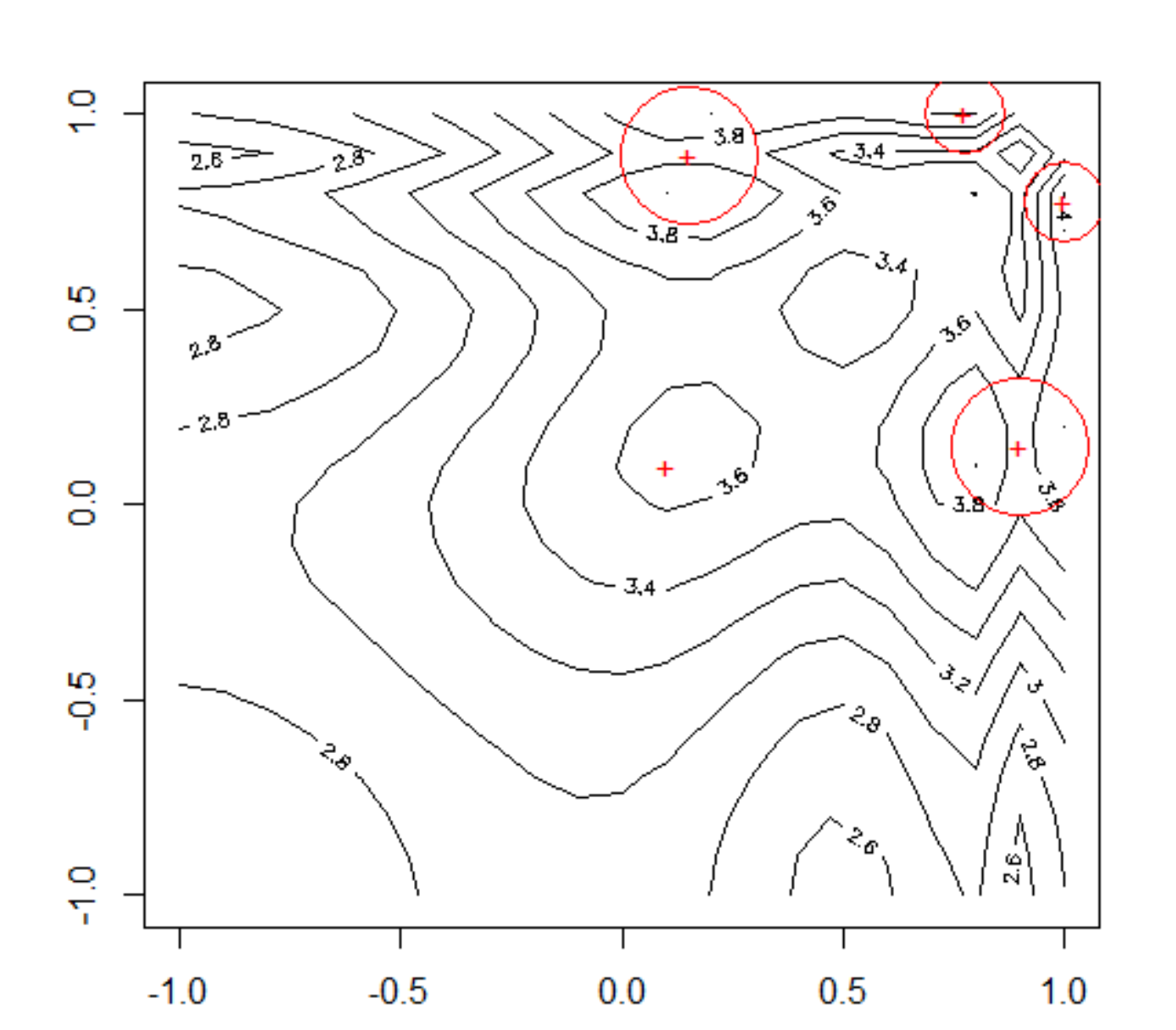}
\includegraphics[viewport = 100 250 500 600, clip, scale = \compscale]{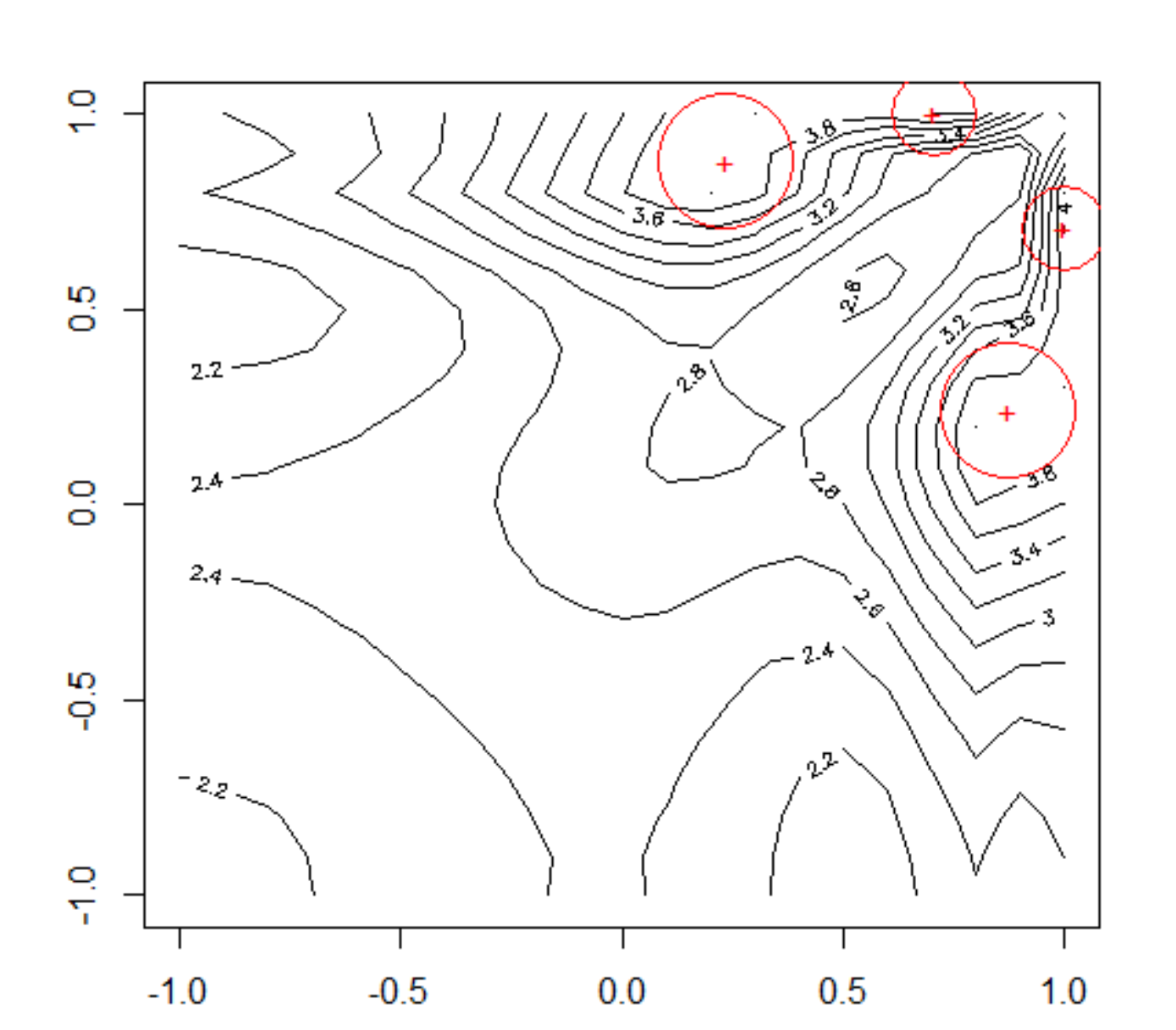}

\includegraphics[viewport = 100 250 500 600, clip, scale = \compscale]{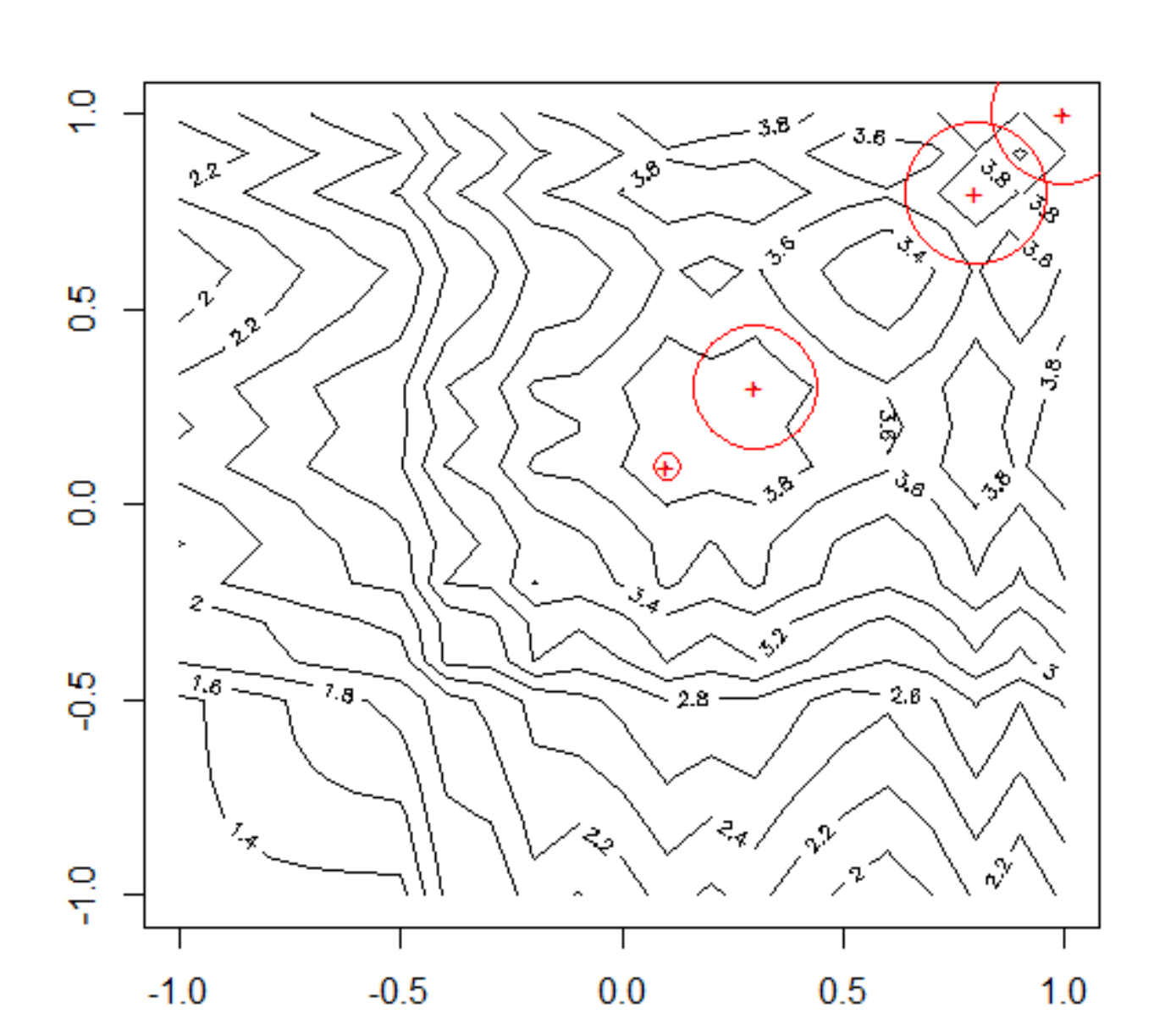}
\includegraphics[viewport = 100 250 500 600, clip, scale = \compscale]{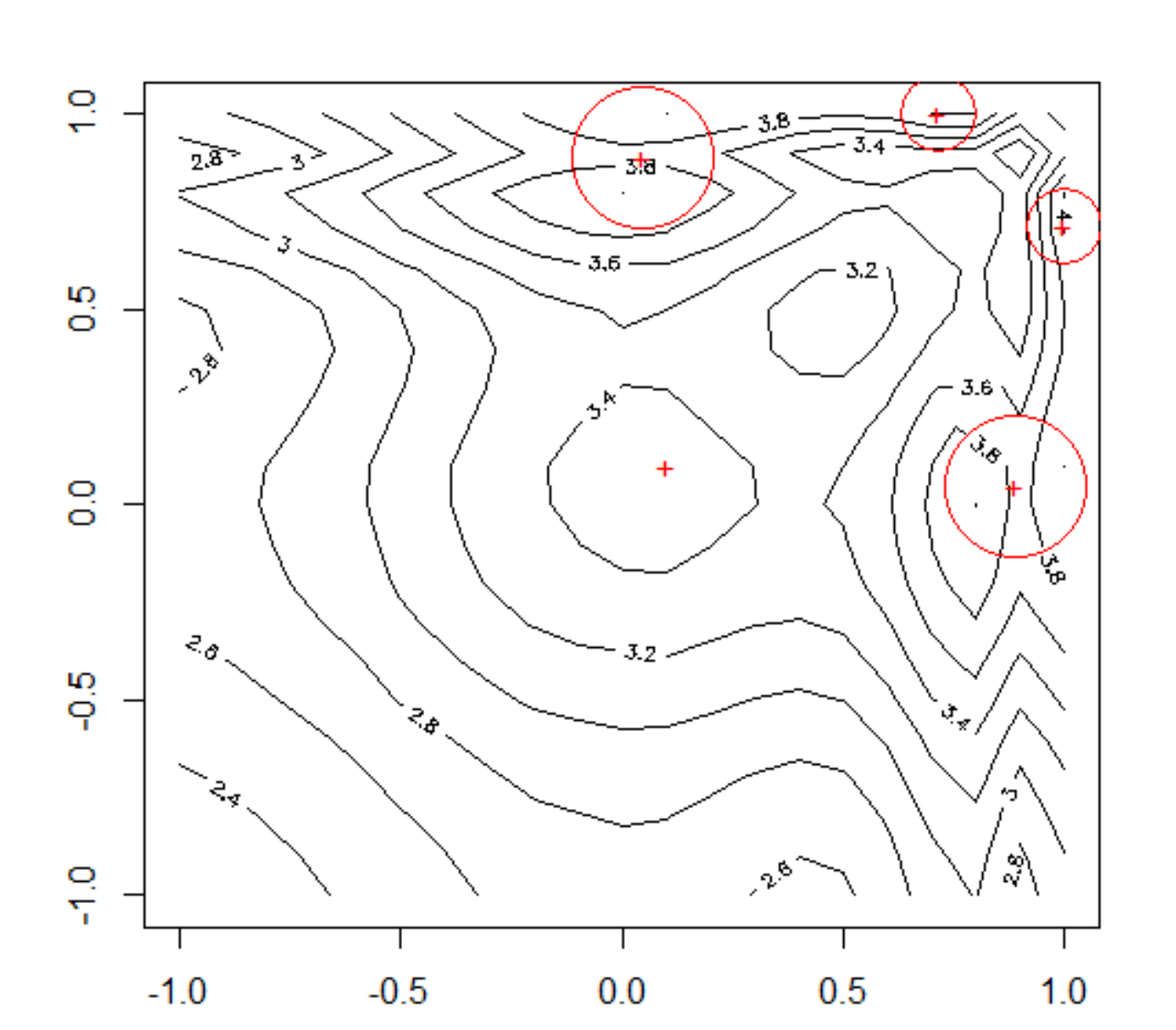}
\includegraphics[viewport = 100 250 500 600, clip, scale = \compscale]{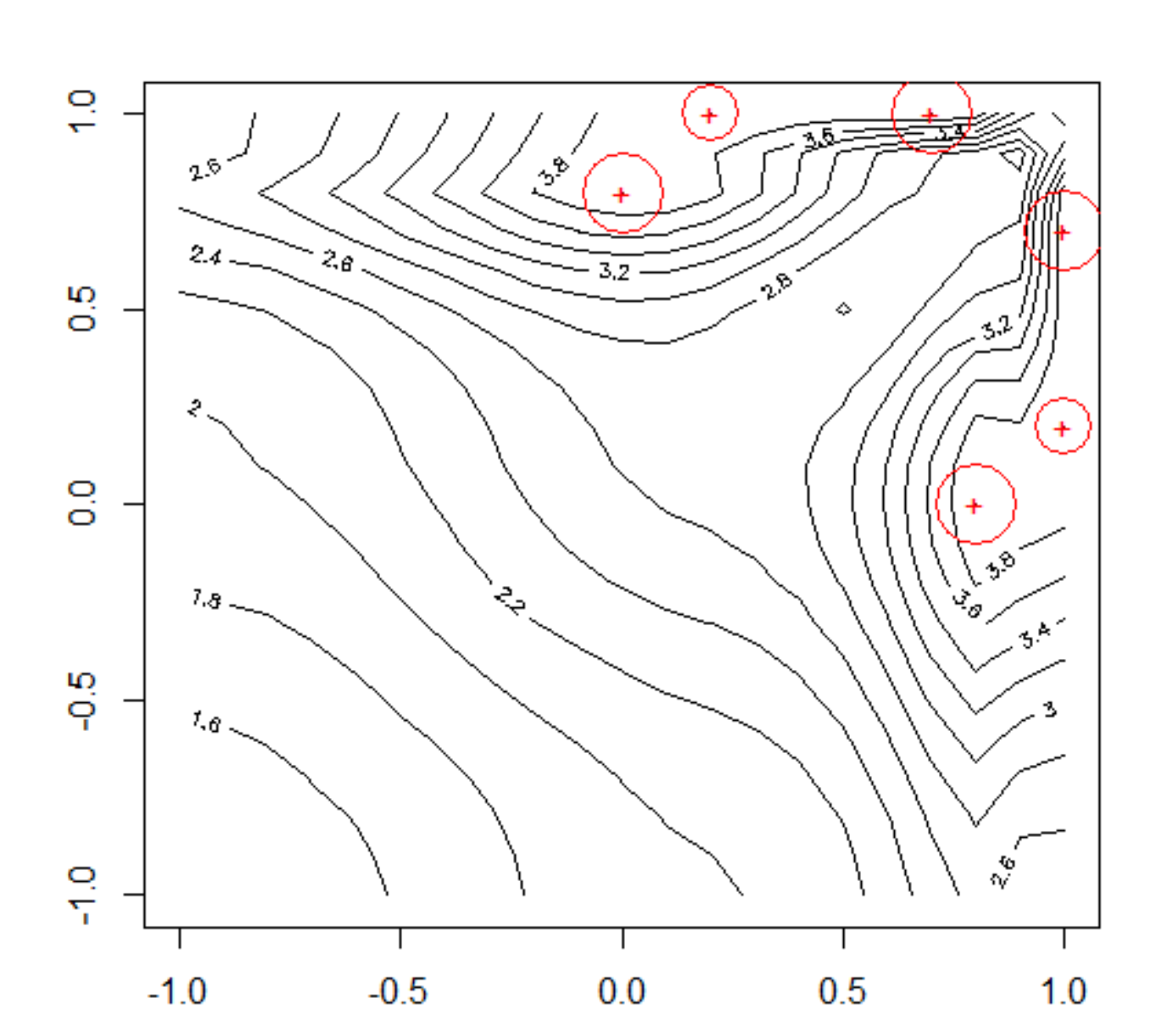}
\end{center}
\caption{\label{fig:toyexample} Optimal designs for the comparative example; rows: Clayton and Gumbel copula; columns levels $\tau_2= \epsilon>0,1/10,1/3$. }
\end{figure}

%

{For reference purposes the optimal design using the independence copula, i.e. an optimal design assuming no block effect, was evaluated. It showed little difference to setting the nominal level for $\tau_2=0$ for a particular copula. In particular the D-efficiencies for the Clayton and Gumbel model were 96.3\% and 99.7\% respectively. This efficiency expectedly decreases as the association within the block increases, for $\tau_2=1/3$ for instance it is already down to 65.0\% and 61.3\% respectively.} 

In \cite{woods+v_11}, robust $D$-optimal designs were found under the same Poisson marginal models and prior distribution but with the dependence described using a GEE approach with an exchangeable correlation matrix and pairwise working correlation of $0.5$. The optimal design found was given by: 
\begin{equation}\label{eq:geedesign}
\xi^\star = \left\{
\begin{array}{ccc}
(.03,1) & (1,.60) & (-.40,.78)\\
.355 & .310 & .335
\end{array}
\right\}\,.
\end{equation}
That is, for example, the first support block is $\zeta_1 = (0.03, 1)$. This design is somewhat different in structure to the copula designs, without the same mirror structure. Quantitatively, the comparison shows the efficiencies under various scenarios given in Table 1. {Surprisingly the design from \cite{woods+v_11} seems to be most compatible with an independence assumption.}
 
\begin{table}[htb]
\begin{tabular}{lllll}
\hline
\multicolumn{1}{l}{Independence} & \multicolumn{1}{l}{Clayton,$\tau_2= \epsilon>0$} & \multicolumn{1}{l}{Clayton,$\tau_2=1/3$} & \multicolumn{1}{l}{Gumbel,$\tau_2= \epsilon>0$} & \multicolumn{1}{l}{Gumbel,$\tau_2= 1/3$} \\ \hline
\multicolumn{1}{c}{ 96.48\%  } & \multicolumn{1}{c}{ 89.85\%  } & \multicolumn{1}{c}{ 84.41\%  } & \multicolumn{1}{c}{ 95.55\%  } & \multicolumn{1}{c}{ 92.96\%  } \\ \hline
\end{tabular}
\caption{D-efficiencies for design~\eqref{eq:geedesign} from the GEE-approach designs under various copula models.} 
\end{table}


\section{Application to the materials example}
\label{sec:application}

In this section we return to the materials testing example to find and assess designs for comparing six materials in block of size two under a variety of modelling assumptions. The measured response is binary, with each material sample either passing or failing a visual check. We label the five novel materials as ``treatments'', with the reference material considered as a control. Marginally, we assume a logistic regression to model the differences between materials set up as
$$
Y_{ij}\sim \mathrm{Bernoulli}\left\{\eta(\mathbf{x}_{ij};\,\boldsymbol{\beta})\right\};\,\quad \eta(\mathbf{x}_{ij};\,\boldsymbol{\beta}) = \mathrm{expit}\left(\beta_0 + \sum_{l=1}^5\beta_ix_{ijl}\right)\,,
$$
where $\mathrm{expit}(u) = 1/\{1 + \exp(-u)\}$, $Y_{ij}$ is the binary response from the $i$th unit in the $j$th block ($i=1,2;\,j=1,\ldots,b$), $\eta(\mathbf{x}_{ij};\,\boldsymbol{\beta}$ is the associated probability of success, $x_{ijl}$ is an indicator variable taking the value 1 if the $i$th unit in the $j$th block was assigned treatment $l$ ($l=1,\ldots,5$) and 0 otherwise, and $\beta_0,\ldots,\beta_5$ are unknown parameters to be estimated. Here, $\beta_0$ is the logit for the reference material, with $\beta_l$ being the difference in expected response, on the logit scale, between the reference material and the $l$th novel material or treatment. 

\begin{figure}[htb] \label{fig:materials} 
\begin{center}
\includegraphics[viewport = 100 250 450 600, clip, scale = .5]{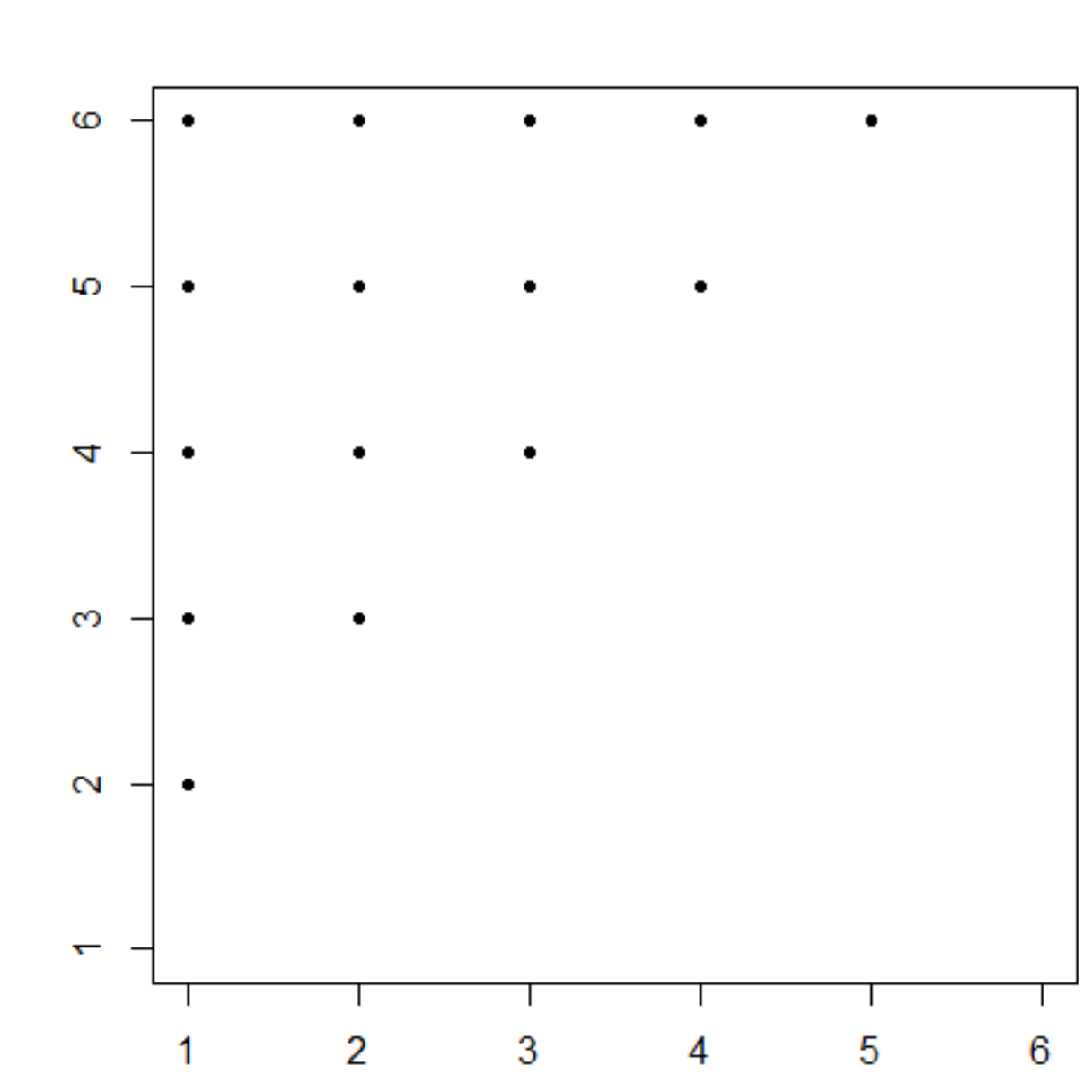} ~~~~~~
\includegraphics[viewport = 100 250 450 600, clip, scale = .5]{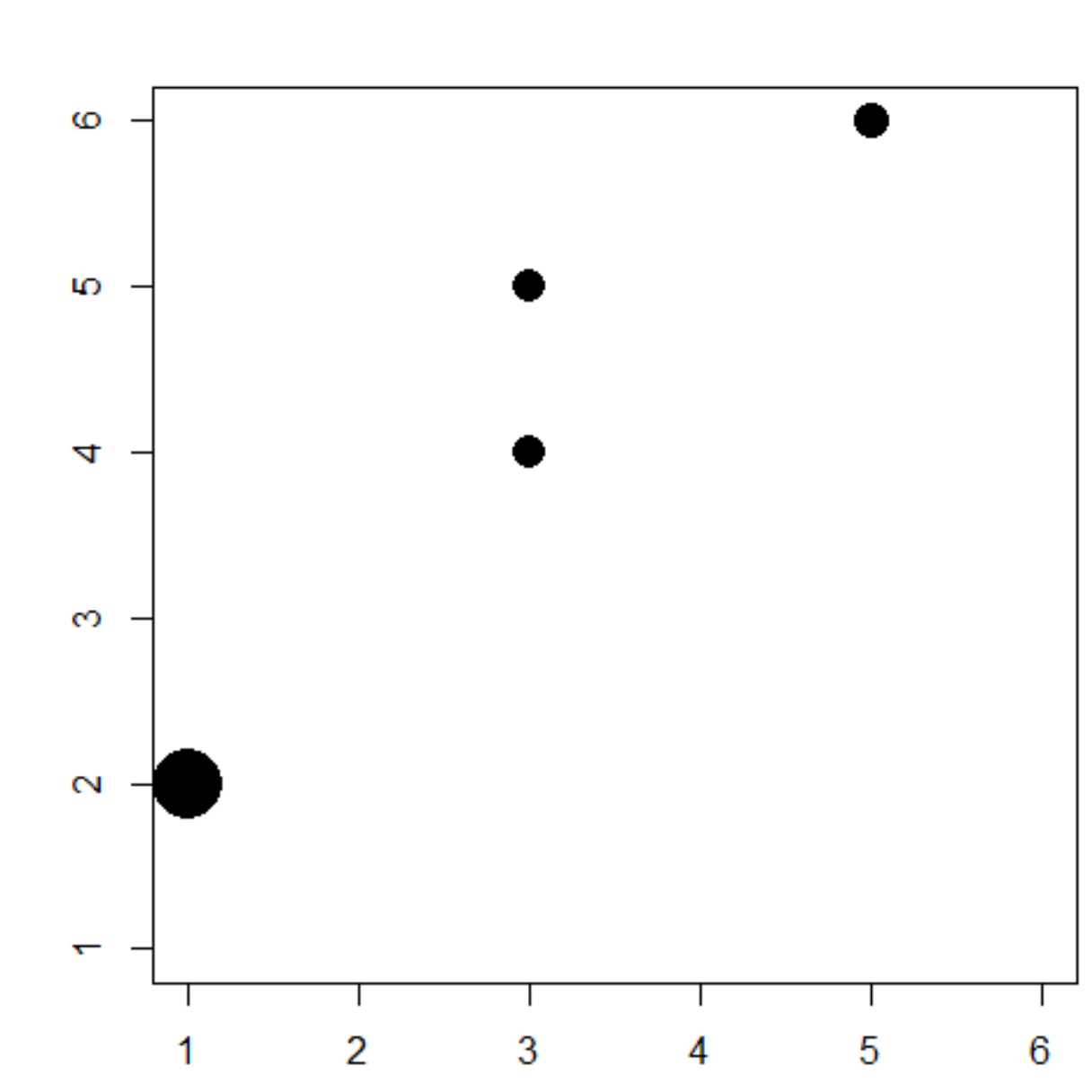}

\includegraphics[viewport = 100 250 450 600, clip, scale = .5]{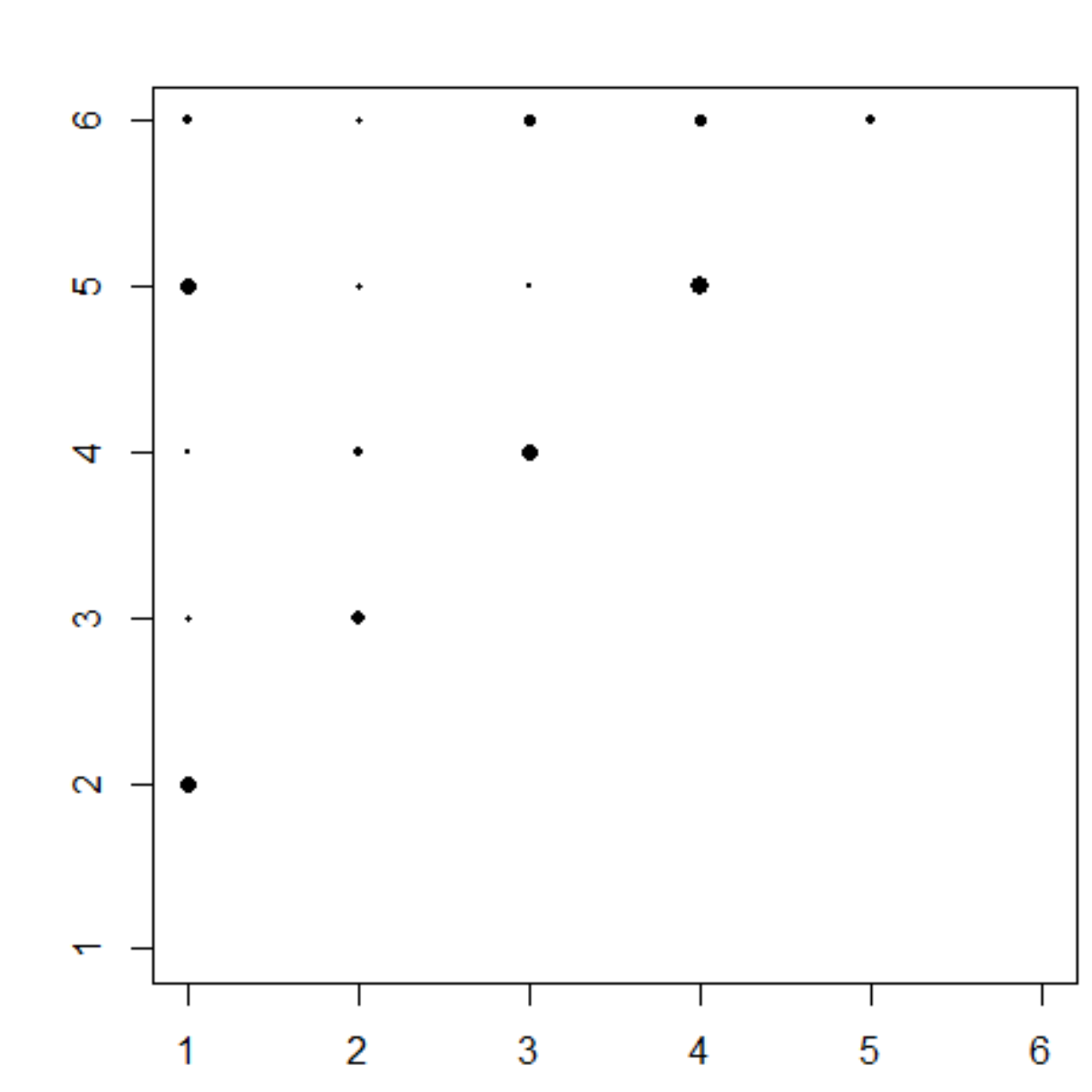} ~~~~~~
\includegraphics[viewport = 100 250 450 600, clip, scale = .5]{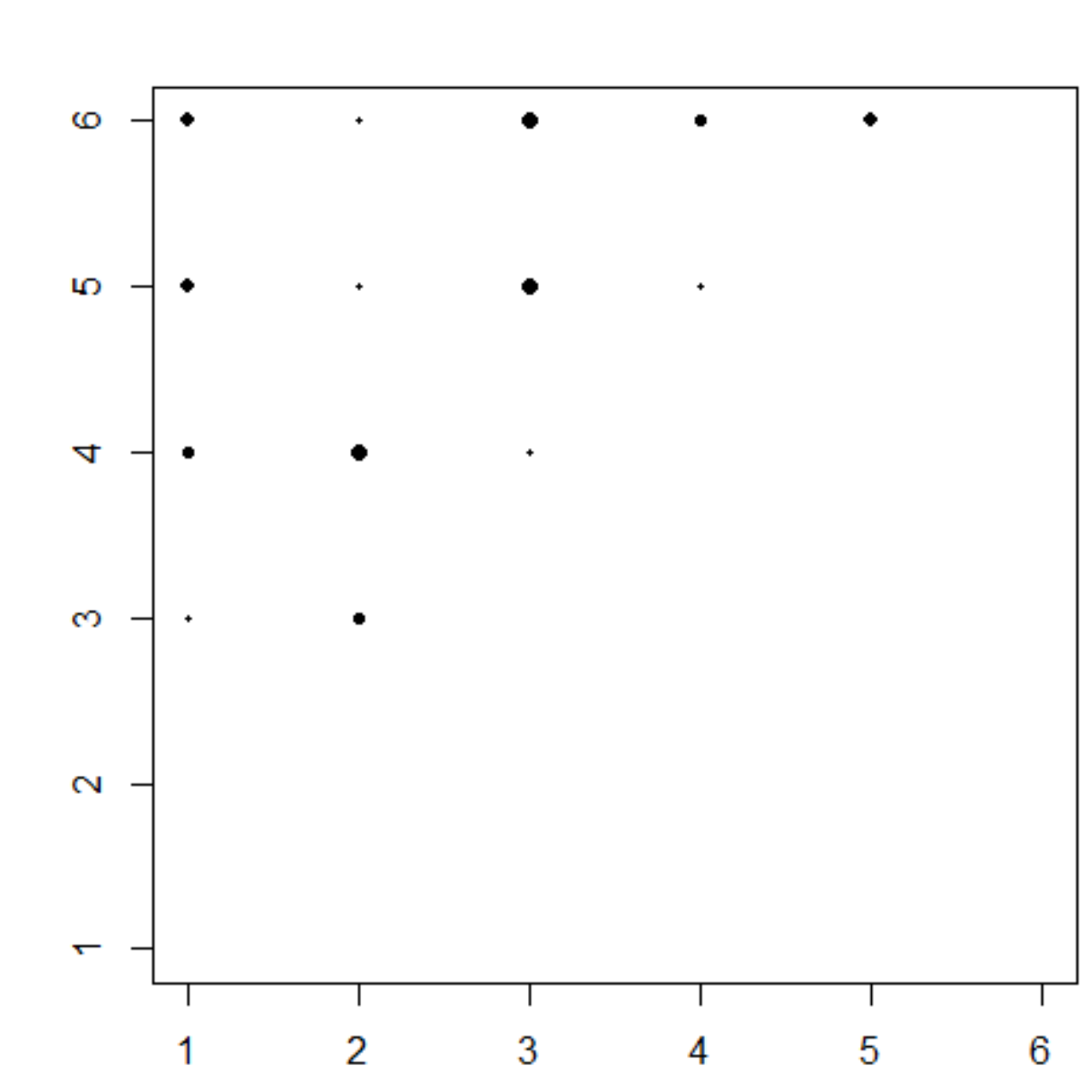}
\end{center}
\caption{Optimal designs for the materials testing example assuming a Gumbel copula with $\tau_2= 0.33$; rows - local and pseudo-Bayesian; columns - assumed parameters or prior mean of $\boldsymbol{\beta}^T=(0,0,0,0,0,0)$ and $\boldsymbol{\beta}^T=(0,-1,2,-3,4,-5)$, respectively. 
}
\end{figure}

The choice of copula and the strength of intra-block association makes little difference to the design selected. However, assuming different marginal models and adopting a local or pseudo-Bayesian approach has a strong impact on the designs. Example designs for the Gumbel copula are shown in Figure~\ref{fig:materials}. 

With a null marginal model, i.e. $\boldsymbol{\beta}^T= (0,0,0,0,0,0)$, when the response variance is constant, the locally D-optimal design contains all material combinations, excluding those blocks containing replicates of a single treatment. This design would also be optimal under a linear model with constant error variance. For different assumed parameter vectors, for example $\boldsymbol{\beta}^T=(0,-1,2,-3,4,-5)$, the optimal design contains only a few distinct treatment and treatment control combinations, with differing weights; here (1,2),(3,4),(4,5) and (5,6) are selected. The (pseudo)-Bayesian approach, assuming a continuous uniform prior on $[-1,1]$ for each $\beta_l$ ($l=0,\ldots,5$) yields designs with unequal weights spread across all material combinations.
Changing to a continuous uniform prior on the space $[-1,1]\times[-2,0]\times[1,3]\times[-4,-2]\times[3,5]\times[-6,-4]$, so centred on $\boldsymbol{\beta}^T=(0,-1,2,-3,4,-5)$, adjusts the weighting of the support blocks to give more emphasis on comparing treatments 2 and 4 and 3 and 5. These pairs of treatments have differences to the control with the same sign. 

\section{Discussion}
\label{sec:disc}

The modeling of block effects by copulas seems a natural choice and allows for elegant separation of the block and the marginal effects. Experimental designs for such models are now readily calculable. 
The pseudo-Bayesian $D_A$-optimality criterion was added to the \texttt{R} package \texttt{docopulae} version 0.4 (see \citealp{docopulae}) with the functions \texttt{wDsensitivity} and \texttt{wDefficiency}, both relying on a prespecified quadrature scheme for evaluation of the integrals. In this paper we have concentrated on finding designs to estimate the complete parameter vector but the implementation provides flexibility for checking for symmetry, model discrimination, etc., as investigated in \citet{perrone+al_17}. 

Our examples are confined to the case $k=2$. Whilst there is no theoretical necessity for that it is difficult to specify high-dimensional parametric copulas with a sufficient range of dependence, for details see the excellent survey of \cite{nikoloulopoulos_13}. However, work on this issue would go well beyond the scope of this paper. It might also be interesting to contrast our findings with some known analytic results for blocks of size two as, for example, given in \cite{cheng} where a Gaussian copula is implicitly assumed.

\section*{Acknowledgements}
We are grateful to Keith Warburton and Rob Ashmore from the UK Defence Science and Technology Laboratory for providing details of the materials testing example. W.G. M{\"u}ller would like to acknowledge the hospitality of the Southampton Statistical Sciences Research Institute during his sabbatical, when this research was initiated. He was partially supported by project grants LIT-2017-4-SEE-001 funded by the Upper Austrian Government, and Austrian Science Fund (FWF): I 3903-N32 and D.C. Woods was partially supported by Fellowship EP/J018317/1 from the UK Engineering and Physical Sciences Research Council.

\renewcommand{\baselinestretch}{1}
\normalsize
\bibliographystyle{asa}
\bibliography{16uk}
\end{document}